\documentclass{article}
\usepackage[accepted]{vietnam} 
\usepackage{natbib}
\usepackage{graphicx}      
\begin{document}
\twocolumn[

\title{Protostellar Jets and Outflows in low-mass star formation}
\author{Masahiro N. Machida}{machida.masahiro.018@m.kyushu-u.ac.jp}
\address{Faculty of Sciences, Kyushu University, Fukuoka 819-0395, Japan}

\vskip 0.5cm 
]

\begin{abstract}

The driving mechanism of protostellar outflows and jets and their effects on the star formation process obtained from 
recent theoretical and numerical studies are described.
Low-velocity outflows are driven by an outer region of the circumstellar disk, while high-velocity jets are driven near an inner edge of the disk. 
The disk angular momentum is effectively transferred by magnetic effects in the outflow and jet driving regions where the magnetic field is well coupled with neutral gas. 
On the other hand, in a high density gas region of the disk (or intermediate region), the magnetic field dissipates and is decoupled from neutral gas.
Thus, in such a magnetically inactive region, no outward flow appears and the disk angular momentum is not effectively transferred by magnetic effects. 
Therefore, in the disk intermediate region, the disk surface density continues to increase and gravitational instability occurs and produce a non-axisymmetric (or spiral) structure.
After spiral arms sufficiently develop, the disk angular momentum is transferred by gravitational torque and a large amount of the disk mass accretes onto the central protostar from the circumstellar disk.
The episodic accretion induces time-variable high-velocity jets.  
The jets do not significantly contribute to a dynamical evolution of the protostar and circumstellar disk, while the low-velocity outflows can eject a large fraction of the infalling gas and determine the final stellar mass. 
\end{abstract}

\section{Introduction}
\vspace{-0.3cm}
Observations in star forming regions indicate that stars form in molecular cloud cores. 
However, angular momentum between molecular cloud cores and stars differs over five orders of magnitude.
Thus, the angular momentum should be removed from gravitationally collapsing cloud cores to produce stars. 
Observations have also implied that protostellar outflows and jets universally appear in the star formation process. 
They are closely related to the angular momentum transfer of the circumstellar disk. 
In addition, these flows are considered to determine the star formation efficiency.
Thus, outflows and jets are very crucial to understand the star formation process. 
Until some years ago, it was considered that only high-velocity jets are driven near a protostar and low-velocity components are entrained by the high-velocity jets in a dense infalling envelope. 
This entrainment scenario had supported by many researchers without clear evidence for many years, although many observed outflows seem not to possess high-velocity components \cite{wu04}. 
To unveil the driving mechanism of these flows and the early star formation process, the evolution of molecular cloud cores until protostars form has been calculated from 1990's using numerical simulations.
Two types of flows were clearly reproduced with a two dimensional ideal magnetohydro-dynamic (MHD) calculation \cite{tomisaka98,tomisaka00,tomisaka02}. 
The driving mechanism of outflows and jets are naturally explained in these studies. 
The outflow is driven by the first core \cite{larson69,masunaga00}, which directly evolves into a circumstellar disk \cite{bate98,machida10}, while the jet is  driven near the protostar \cite{machida06,machida08,machida14b}.

Recently, high-velocity jets and low-velocity outflows were also reproduced in three dimensional radiation non-ideal  MHD simulations \cite{tomida13,tomida15,tsukamoto15a,tsukamoto15b}.
A long-term evolution of outflows and jets were also studied  in recent simulations without solving the radiative transfer equation \cite{machida12,machida13,machida14b}. 
These studies confirmed that properties of outflows and jets taken from numerical simulations are well agreement with observations. 
In this article, the recent development of theoretical and numerical studies for protostellar outflows and jets is presented.

\begin{figure*}
\centering
$\begin{array}{cc}
\includegraphics[angle=0,width=14.5cm]{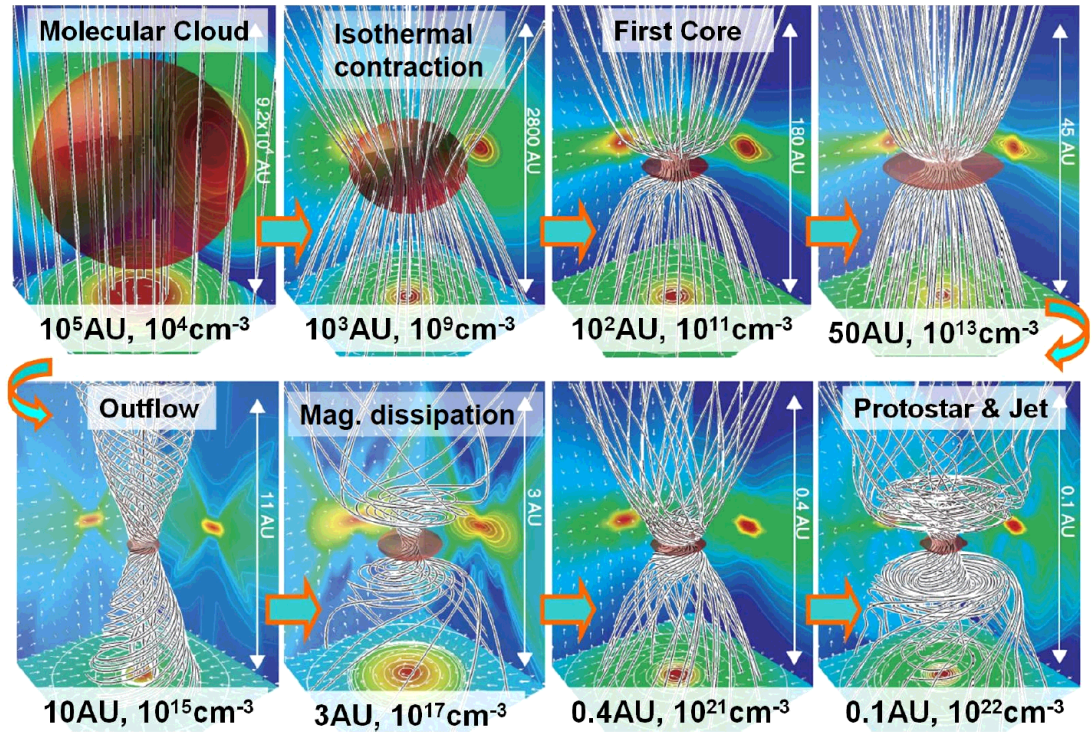} 
\end{array}$
\caption{ 
Cloud evolution in a gas collapsing phase. 
The central red region in each panel corresponds to a high-density gas region.
In each panel, white lines are magnetic field lines.
Density distributions and velocity vectors are projected on each wall surface. 
Relevant phenomena at each epoch is described in the upper side of each panel.
The spatial scale and central density are described in the lower side of each panel.
}
\label{fig:1}
\vspace{-0.5cm}
\end{figure*}

\section{Types of Numerical Simulations} 
\vspace{-0.3cm}
The simulations of protostellar jets and outflows are roughly classified into three types \cite{Konigl00,shang07}.
The first one starts the calculation of the jet driving after a protostar sufficiently evolves, in which properties of the circumstellar disk, magnetic field lines, density and velocity distributions are artificially given, and many parameters are adjusted to drive jets. 
In this type of simulations, artificial initial settings and highly adjusted parameters determine the result. 
Thus, although a necessary condition for driving a jet might be investigated, we cannot know whether such highly artificial conditions and parameters are realistic in the star formation process. 
The second one is the jet propagation simulation, in which jets are artificially injected into a computational domain to investigate the interaction between the jet and infalling gas. 
Thus, the driving mechanism of jets cannot be identified in such simulations.

The last one is the cloud collapse and core accretion simulation, in which the calculation stars from prestellar cloud core stage, and the cloud evolution is calculated before and after the protostar formation. 
In this type of simulations, since only an initial prestellar cloud core is set, the circumstellar disk, outflows and jets naturally appear. 
Thus, we can better understand various phenomena appeared in the star formation process  \cite{tomisaka02}.
Fig.~\ref{fig:1} shows an example of such a cloud collapse simulation \cite{machida04,machida05a}, in which the cloud evolution from prestellar stage until the protostar formation is presented. 
The cloud collapse simulation needs to cover a wide range of spatial scales ($\sim0.01-10^5$\,AU) and density contrasts ($\sim10^4-10^{22}\,{\rm cm}^{-3} $) using special techniques such as nested grid and adaptive mesh refinement methods \cite{machida05a,machida05b,machida10}.
In the below, the driving mechanism of outflows and jets are explained according to recent results of the cloud collapse simulations.

\section{Low-velocity Outflow} 
\vspace{-0.3cm}
Firstly, the cloud evolution and protostar formation in collapsing clouds are described (Fig.~\ref{fig:1}). 
The cloud contracts isothermally until its central density exceeds $\sim10^{10}\,{\rm cm}^{-3} $.
Then, the central region becomes optically thick against the dust thermal emission, and the first adiabatic core (or first core) forms \cite{larson69}. 
The contraction timescale is shorter than the rotational timescale in the first core, and magnetic field lines are twisted and inclined against the rotation axis around the first core. 
Therefore, the first core drives outflows, which transfer the angular momentum,  mainly by the magnetocentrifugal mechanism \cite{blandford92}. 

On the other hand, since the ionization rate inside the first core (or in the region of $10^{12} < n < 10^{15}\,{\rm cm}^{-3} $) is significantly low, the magnetic field dissipates mainly by Ohmic dissipation. 
In addition to mass accretion, the removal of angular momentum and magnetic field from the first core causes a gradual increase of the central density and temperature.
Then, molecular hydrogens begin to be dissociated when the central density reaches $n \sim 10^{15}\,{\rm cm}^{-3} $ (or $T \sim 2000$\,K), and the second collapse occurs \cite{larson69}. 
Thereafter, the dissociation of molecular hydrogens is completed, and a protostar forms when the central density reaches $n\sim10^{20}-10^{22}\,{\rm cm}^{-3} $. 
The mass and radius of the protostar at its formation epoch are $M_{\rm ps}\sim10^{-3}\thinspace M_\odot$ and $r_{\rm ps}\sim0.01$\,AU, respectively \citep{larson69,masunaga00}.

\begin{figure*}
\vskip -0.5cm
\centering
$\begin{array}{cc}
\includegraphics[angle=0,width=16.5cm]{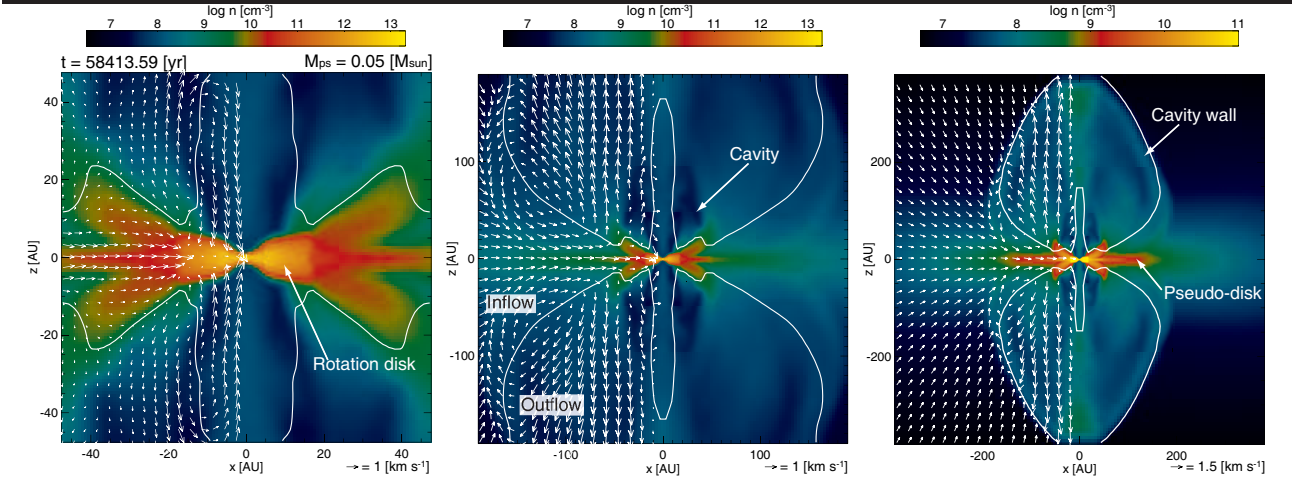} 
\end{array}$
\caption{
Density distributions (color) and velocity vectors (arrows; left side of each panel) at the same epoch.
The spatial scale differs in each panel. 
White lines correspond to the boundary between infall and outflow motions.
The elapsed time after the cloud collapse begins $t$ and the protostellar mass $M_{\rm ps}$ are descried.
}
\label{fig:2}
\vspace{-0.5cm}
\end{figure*}

Only a central part of the first core collapses to form a protostar, while the remainder evolves into a circumstellar or Keplerian disk. 
Thus, the low-velocity outflow, which is driven by the first core before the protostar formation, is driven by the circumstellar disk or remnant of the first core after the protostar formation. 
Since the transition of the first core (remnant) to the Keplerian disk is smooth, the low-velocity outflow continues to appear even after the protostar formation \cite{machida13}. 
As shown in Fig.~\ref{fig:2}, low-velocity magnetic winds emerge only from the outer
 disk region because the magnetic field dissipates and weakens in the inner disk region. 
Thus, the launching region of low-velocity outflows is limited in the range of $\sim1-100$\,AU in the early phase of the star formation.
The inner boundary of the outflow launching region ($>1$\,AU) is determined by the magnetic dissipation region, while the outer boundary corresponds to the outer edge of the Keplerian rotating disk. 
Both the inner and outer boundaries vary with time. 
In the inner disk region, no low-velocity outflow appears and the infall motion is observed as seen in the left panel of Fig.~\ref{fig:2}. 

\section{High-velocity Jet} 
\vspace{-0.3cm}
The magnetic field is again well coupled with neutral gas near the protostar because the gas temperature and ionization rate become high. 
Thus, magnetic field lines are twisted and amplified by the rotation of the protostar and disk inner region. 
Therefore, another magnetically driven winds emerge near the protostar. 
A gravitational potential well near the protostar is deep, and the wind velocity corresponds to the escape velocity from the protostar.
Thus, the maximum speed of the flows near the protostar is as high as $\sim100\,{\rm km\, s}^{-1}$. 
In addition, since the magnetic field in the circumstellar disk dissipates, the field around the protostar is very weak before the amplification due to the rotation motion. 
Thus, the magnetic field lines are strongly twisted, and the toroidal component dominates the poloidal component.
Therefore, the high-speed flow is driven mainly by the magnetic pressure gradient force and the toroidal field collimates the flow by hoop stress \cite{tomisaka02,machida08}.

\begin{figure*}
\centering
$\begin{array}{cc}
\includegraphics[angle=0,width=16.5cm]{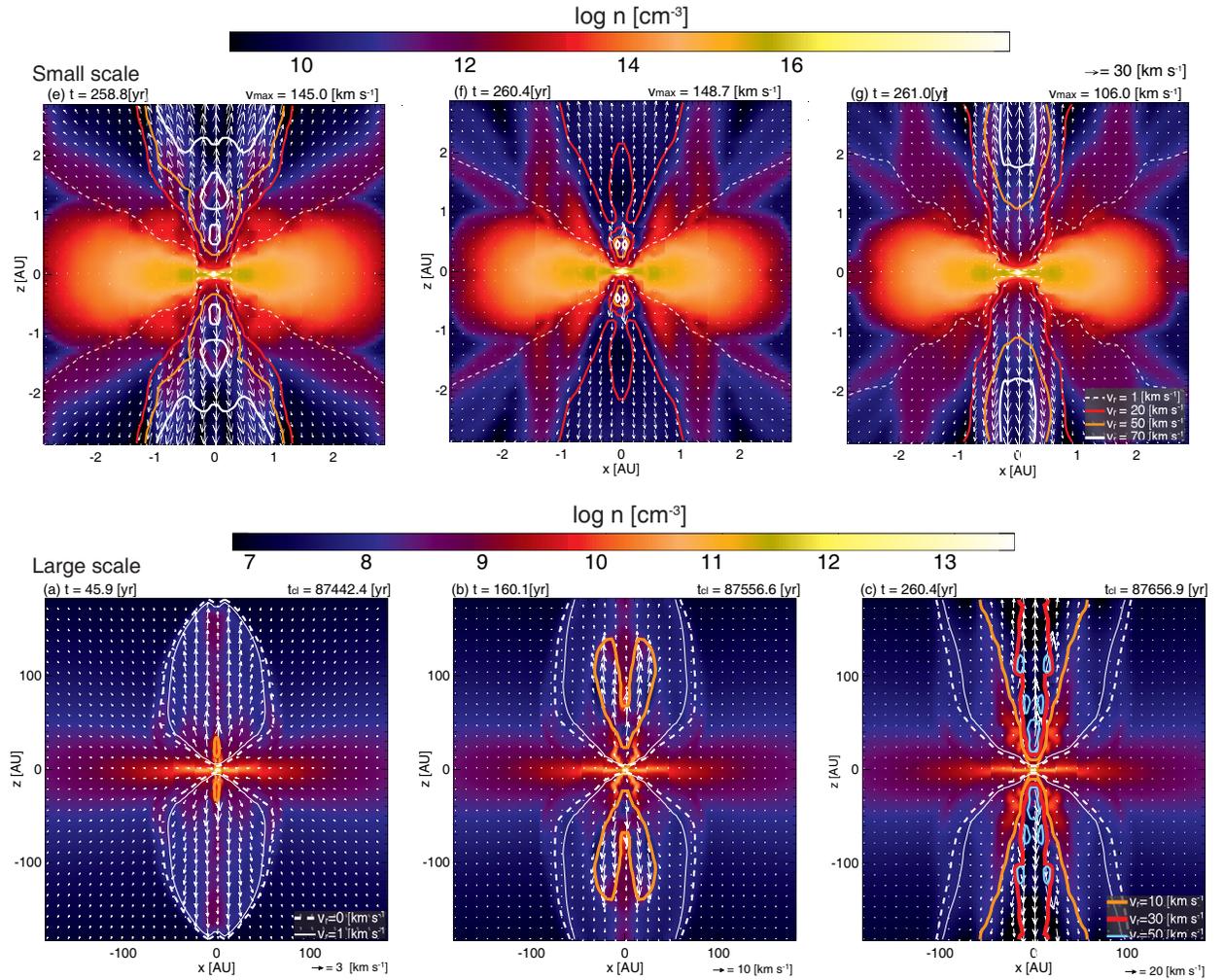} 
\end{array}$
\caption{
Cloud evolution in the mass accretion phase after a protostar formation.
Spatial scales differ in upper and lower panels.
The density distribution (color) and velocity vectors (arrows) are plotted in each panel. 
The contours corresponds to the iso-velocity of outflow and jet \cite{machida14b}. 
The elapsed time after the protostar formation $t$, the elapsed time after the cloud collapse begins and the maximum jet velocity $v_{\rm max}$ are described in each panel.
}
\label{fig:3}
\vspace{-0.5cm}
\end{figure*}

\begin{figure*}
\centering
$\begin{array}{cc}
\includegraphics[angle=0,width=16.5cm]{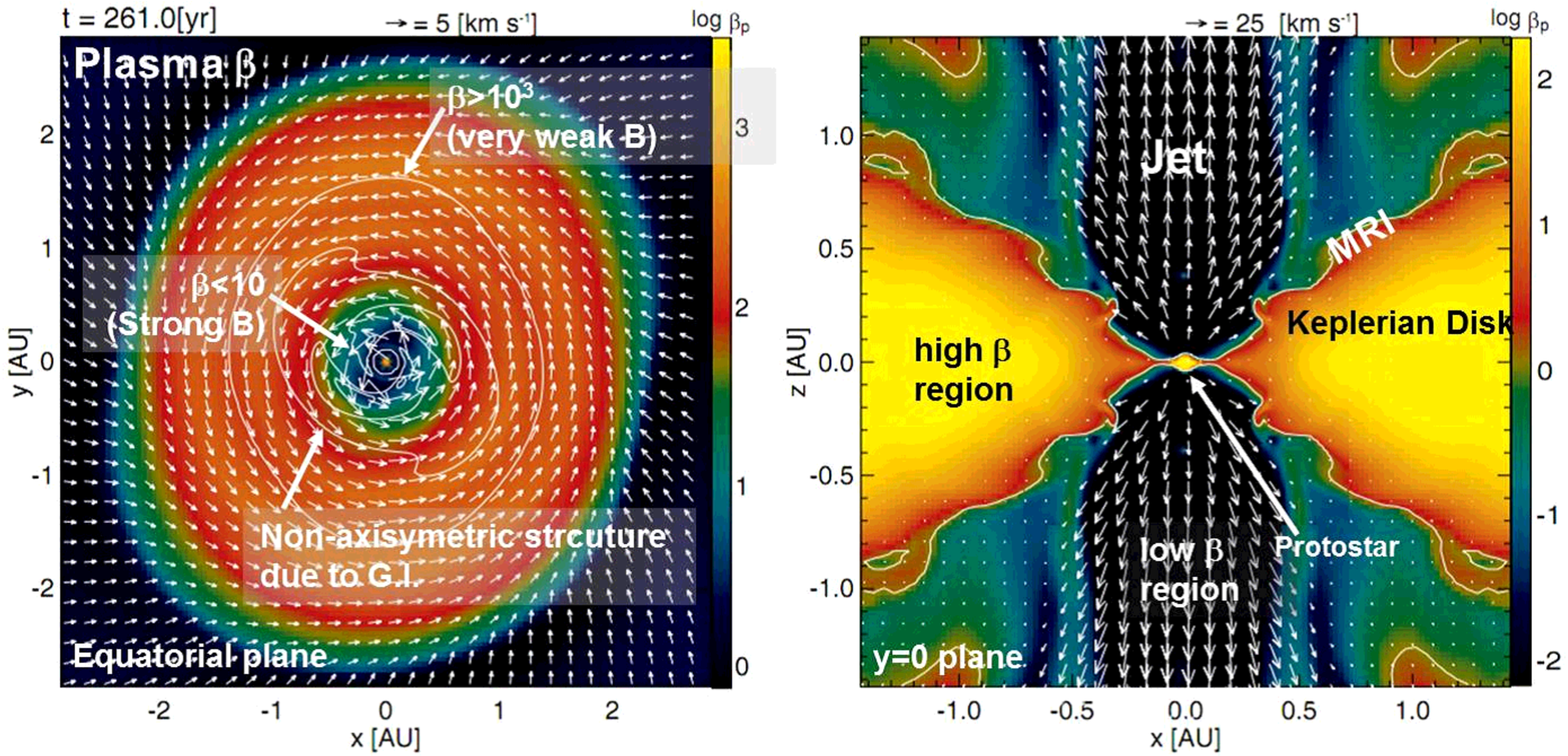} 
\end{array}$
\caption{
Plasma beta (color)  and velocity vectors on the equatorial plane (left) and $y=0$ plane (right) in the mass accretion phase \cite{machida14b}.
The density contour is overplotted by white contour lines. 
}
\label{fig:4}
\vspace{-0.5cm}
\end{figure*}

As shown in Fig.~\ref{fig:3} upper panels, protostellar jets shows a time variability. 
The time variability is caused by the episodic accretion which attributes to the gravitational instability in a magnetically inactive region of circumstellar disk.
Fig.~\ref{fig:4} shows the plasma beta $\beta$ on the equatorial plane (left) and $y=0$ plane (right) at a very early phase of star formation. 
On the equatorial plane, the plasma beta exceeds $\beta >100$ in the range of $0.5\,{\rm AU}<r<2\,{\rm AU}$ where the magnetic field dissipates (i.e. magnetically inactive region).
Thus, the disk angular momentum is not effectively transferred by magnetic effects \cite{machida11,machida14a}. 
Therefore, the disk surface density becomes high and gravitational instability (GI) occurs \cite{toomre64}, and thus non-axisymmetric or spiral structures develop due to GI.
Then, the angular momentum in this region is effectively transferred by the gravitational torque, and a large amount of the disk mass accretes onto the central protostar. 
After the violent mass accretion, the disk surface density decreases and the disk recovers a stable state. 
Thereafter, the mass accretion from the infalling envelope onto the disk increases the disk surface density and GI is again induced. 
In this way, in the circumstellar disk, the transition from gravitationally stable to unstable states is cycled, and the mass accretion episodically occurs.
This episodic mass accretion causes a time-variable mass ejection (or high-velocity jet), as seen in Fig.~\ref{fig:3} upper panels. 

\section{Driving Mechanism and Configuration of Jets and Outflows} 
As described above, two different flows (low-velocity outflows and high-velocity jets) appear in the star formation process. 
The low-velocity outflow are driven by the disk outer region, while the jets are driven near the protostar (or near the inner edge of the circumstellar disk).  
In the intermediate region of the circumstellar disk, no outward flow appear because the dissipation of magnetic field occurs. 
The velocity difference in low and high-velocity flows attributes to the difference in their launching points.
The speed of these flows roughly correspond to the escape velocity at their launching points.

The outflows have a wide opening angle, while the jets have a very good collimation.
The difference shape of these flows is caused by their driving mechanism. 
The magnetocentrifugal mechanism is effective for the low-velocity outflow driven by the outer disk region where the magnetic field is strong.
On the other hand, the magnetic pressure driven mechanism is effective for the high-velocity jet due to weak ambient magnetic field \cite{machida08}.

\begin{figure*}
\centering
$\begin{array}{cc}
\includegraphics[angle=0,width=15.5cm]{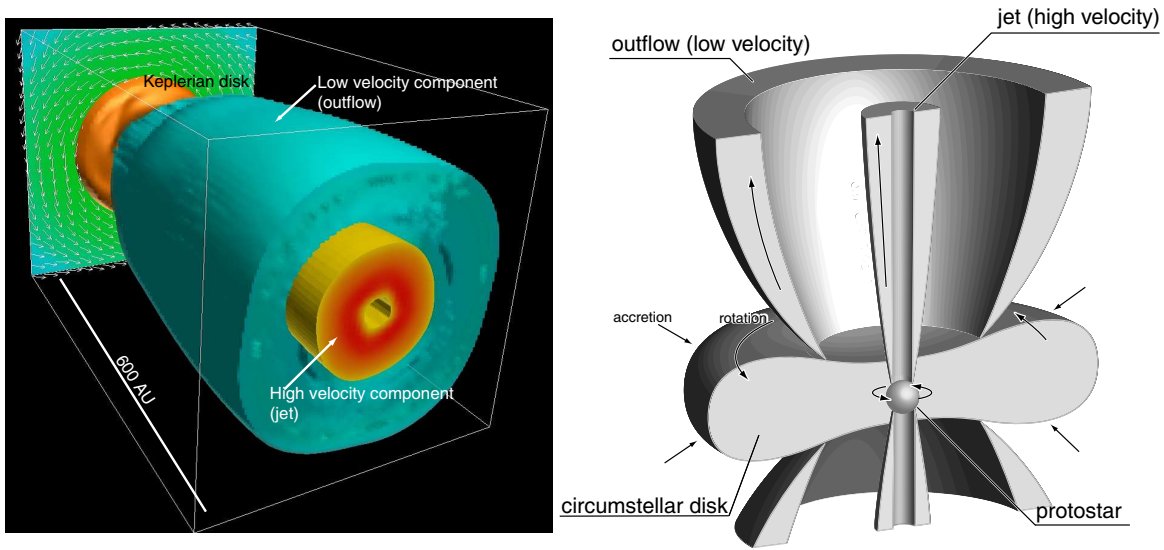} 
\end{array}$
\caption{
Three dimensional view of outflow, jet and circumstellar disk in the mass accretion phase taken from simulation (left). 
Schematic view around a protostar (right).
}
\label{fig:5}
\vspace{-0.5cm}
\end{figure*}

The outflow appears before the protostar formation and  jet emergence.
Thus, high-velocity jets propagate into low-velocity outflow as seen in Fig.~\ref{fig:3} lower panels. 
In other words, high-velocity components are enclosed by low-velocity components, as shown in Fig.~\ref{fig:5}. 
The outflows are steadily driven by the outer disk region, while the jet intermittently appear near the protostar. 
This is because the steady accretion is realized in the (low-velocity) outflow driving region and the episodic accretion occurs in the jet driving region.  
In addition, jets sometimes undergoes a (plasma) kink instability and disappear because the toroidal field dominates in jets \cite{machida08}. 
Thus, it is natural that only a low-velocity outflow is observed without high-velocity components around a protostar. 
The cloud evolution and core accretion simulations could explain the prestellar outflow associated with/without high-velocity jets.

\section{Summary}
\vspace{-0.3cm}
After the discovery of a protostellar outflow, researchers had been discussed the driving mechanism of low-velocity outflows and high-velocity jets. 
Many researchers believed that the low-velocity component is entrained by high-velocity jets that are driven directly by the inner disk region (or inner edge of the disk). 
Now, this entrainment hypothesis is rejected by recent high-spatial resolution observation \cite{bjerkeli16}.

On the other hand, recent cloud collapse simulations have unveiled the driving mechanism of jets and outflows.
The low-velocity outflow is originated from the first core. 
The first core (remnant) becomes to the circumstellar disk after the protostar formation, and the magnetic field dissipates in the inner region of circumstellar disk.
Thus, the low-velocity outflows are driven by the outer region of the circumstellar disk where the magnetic field is well coupled with neutral gas. 
Inside the disk, since the magnetic fields dissipates and the angular momentum is not effectively transferred by magnetic effects, the gravitational instability occurs and the episodic mass accretion onto the protostar is induced. 
The magnetic field is well coupled with neutral gas around the protostar and the episodic accretion causes the episodic jets.

The difference in the flow speeds between outflows and jets is attributed to the difference in outflow and jet driving regions, because the flow speed corresponds to the escape velocity. 
On the other hand, the difference in the degree of collimation (wide-angle outflow and well-collimated jet) is caused by their primary driving mechanism (magnetocentrifugal or magnetic pressure gradient driving mechanisms).
In this way, the driving mechanism of protostellar outflows and jets, and their effects on the star formation are well explained by the cloud collapse simulations.
In addition, their results show good agreements with recent ALMA observations \cite{tomida16}.

However, the cloud collapse simulations have not yet arrived at the planet formation stage. 
The planet formation is considered to begin in the circumstellar disk after the mass accretion onto the protostar significantly decreases. 
To clarify the planet formation and bridge the gap between the star and planet formation, we need to proceed the cloud collapse simulation until a further evolutional stage.

\section*{Acknowledgments}
\vspace{-0.3cm}
This research used computational resources of the HPCI system provided by (Cyber Sciencecenter, Tohoku University; Cybermedia Center, Osaka University, Earth Simulator, JAMSTEC) through the HPCI System Research Project (Project ID:hp150092, hp160079).
This work was supported by Grants-in-Aid from MEXT (25400232, 26103707).

\end{document}